\documentclass[iop]{emulateapj}
\usepackage{stackengine}
\usepackage{natbib}
\shorttitle{Isotopic Ratios of Carbon and Oxygen in Titan's CO using ALMA}
\shortauthors{Serigano et al.}

\submitted{Accepted for publication in ApJ Letters}

\begin{document}

\title{Isotopic Ratios of Carbon and Oxygen in Titan's CO using ALMA}

\email{jserigano4@jhu.edu}
\author{Joseph Serigano IV\altaffilmark{2,1,*}, C. A. Nixon\altaffilmark{1}, M. A. Cordiner\altaffilmark{2,1}, P. G. J. Irwin\altaffilmark{3}, N. A. Teanby\altaffilmark{4}, S. B. Charnley\altaffilmark{1}, J. E. Lindberg\altaffilmark{1}}
\altaffiltext{*}{Current affiliation: Department of Earth and Planetary Sciences, Johns Hopkins University, Baltimore, MD 21218}
\altaffiltext{1}{NASA Goddard Space Flight Center, 8800 Greenbelt Road, Greenbelt, MD 20771, USA.}
\altaffiltext{2}{Department of Physics, Catholic University of America, Washington, DC 20064, USA.}
\altaffiltext{3}{Atmospheric, Oceanic, and Planetary Physics, Clarendon Laboratory, University of Oxford, Parks Road, Oxford, OX13PU, UK}
\altaffiltext{4}{School of Earth Sciences, University of Bristol, Wills Memorial Building, Queens Road, Bristol, BS8 1RJ, UK.}

\begin{abstract}

We report interferometric observations of carbon monoxide (CO) and its isotopologues in Titan's atmosphere using the Atacama Large Millimeter/submillimeter Array (ALMA). The following transitions were detected: CO (J~= 1--0, 2--1, 3--2, 6--5), $^{13}$CO (J = 2--1, 3--2, 6--5), C$^{18}$O (J = 2--1, 3--2), and C$^{17}$O (J = 3--2). Molecular abundances and the vertical atmospheric temperature profile were derived by modeling the observed emission line profiles using NEMESIS, a line-by-line radiative transfer code. We present the first spectroscopic detection of $^{17}$O in the outer solar system with C$^{17}$O detected at \textgreater 8$\sigma$ confidence. The abundance of CO was determined to be 49.6 ${\pm}$ 1.8 ppm, assumed to be constant with altitude, with isotopic ratios $^{12}$C/$^{13}$C = 89.9 ${\pm}$ 3.4, $^{16}$O/$^{18}$O = 486 ${\pm}$ 22, and $^{16}$O/$^{17}$O = 2917~${\pm}$~359. The measurements of $^{12}$C/$^{13}$C and $^{16}$O/$^{18}$O ratios are the most precise values obtained in Titan's atmospheric CO to date. Our results are in good agreement with previous studies and suggest no significant deviations from standard terrestrial isotopic ratios. 
\end{abstract}


\section{Introduction}

Titan, Saturn's largest moon, exhibits strong emission features in the submillimeter region, where the rotational lines of small, polar molecules are prominent and allow us to probe the vertical structure of the atmosphere. In the dense, nitrogen-dominated atmosphere of Titan, photodissociation of molecular nitrogen and methane leads to a wealth of complex hydrocarbons and nitriles in small abundances, permitting the study of atmospheric processes that may be similar to the reducing atmosphere of the early Earth. Additionally, the existence of oxygen on Titan facilitates the synthesis of molecules of potential astrobiological importance. 

Carbon monoxide, the fourth most abundant molecule in Titan's atmosphere, was first detected by \citet{lutz83} in the near-infrared using the 4-m Mayall telescope at Kitt Peak National Observatory. The existence of atmospheric CO led to many questions about its origins. Early investigations \citep{yung84, toublanc95, lara96, wong02, baines06, wilson04} assumed primordial origins, volcanic outgassing, and/or external sources from H$_{2}$O via micrometeorite ablation but failed to reproduce the observed abundances of CO.

A viable solution regarding the source of atmospheric CO was realized with the Cassini Plasma Spectrometer (CAPS) detection of O$^{+}$ precipitating into Titan's upper atmosphere at an influx rate of $\sim$10$^{6}$ cm$^{2}$ s$^{-1}$\citep{hartle06}. \citet{horst08} provided a photochemical model in which the abundances of CO, CO$_{2}$, and H$_{2}$O are all reproduced as expected using an oxygen flux in agreement with the CAPS detection and an OH flux in agreement with predicted production from micrometeorite ablation. In this model, reactions between neutral ground-state O atoms (O($^{3}$P)) and CH$_{3}$ can accurately reproduce the CO abundance, ruling out the necessity to include outgassing from Titan's interior or primordial remnants as a source of CO. More recently, \citet{dobri14} presented a model with a more complex oxygen chemical scheme, incorporating a coupling between hydrocarbon, oxygen, and nitrogen chemistries. However, their model, which uses recent constraints on the lower atmospheric abundance of water vapor \citep{cottini12,moreno12} and considers multiple scenarios for the origin of oxygen, is still unable to definitively differentiate the source of CO as internal or external. 

\begin{deluxetable}{lll}
\tablewidth{0pt}
\tablecaption{Recent Measurements of the CO Abundance in Titan's Atmosphere\label{tbl-1}}
\tablehead{
\colhead{Instrument} &
\colhead{CO VMR (ppm)} &
\colhead{Reference}}
Cassini CIRS & 47${\pm}$8 & \citet{dekok07}\\
Cassini CIRS & 55${\pm}$6 & \citet{teanby10}\\
APEX SHeFI &30${\pm}$${15 \atop 8}$ & \citet{rengel11}\\
Herschel SPIRE & 40${\pm}$5 & \citet{courtin11}\\
SMA & 51${\pm}$4 & \citet{gurwell11}\\
IRTF CSHELL & 51${\pm}$7 & \citet{debergh12}\\
Herschel SPIRE & 47${\pm}$7 & \citet{teanby13}\\
Herschel PACS & 50${\pm}$2  & \citet{rengel14}\\
\end{deluxetable}

\begin{deluxetable*}{cccccccccc}
\centering
\tablecaption{Observational Parameters of CO Lines}
\tablehead{
\colhead{Isotopologue} &
\colhead{\stackanchor{Rest}{Freq.~(GHz)}} &
\colhead{\stackanchor{Observation}{Date}} &
\colhead{\stackanchor{Integration}{Time (s)}}  &
\colhead{\stackanchor{\# of}{Antennae}} &
\colhead{\stackanchor{Spectral}{Res.~(kHz)\tablenotemark{a}}} &
\colhead{\stackanchor{Spatial}{Res.\tablenotemark{b}}} &
\colhead{\stackanchor{Altitude}{Sensitivity (km)\tablenotemark{c}}} &
\colhead{Project Code}}
\startdata
CO (1--0) & 115.271 & 2012 Aug 26 & 149 & 23 & 976 & $1.87''\times1.30''$  & 140 - 310 & 2011.0.00772.S\\ 
CO (2--1)  & 230.538 & 2012 Aug 09 & 231 & 23 & 1953& $0.99''\times0.66''$ &  200 - 340 & 2011.0.00319.S\\ 
 &  & 2014 Apr 04 & 158 & 34 & 1953& $0.87''\times0.72''$ &  200 - 340 & 2012.1.00261.S\\
CO (3--2)  & 345.796 & 2012 Aug 25 & 228 & 28 & 976& $0.66''\times0.46''$ &  320 - 500 & 2011.0.00419.S\\ 
CO (6--5)  & 691.473 & 2012 Jun 05 & 230 & 21 & 1953& $0.35''\times0.28''$ & 420 - 550 & 2011.0.00724.S\\ 
\\[.1mm]
$^{13}$CO (2--1)  & 220.399 & 2012 Jun 15 & 230 & 20 & 122& $1.19''\times0.79''$ &  150 - 270 & 2011.0.00465.S\\ 
$^{13}$CO (3--2)  & 330.588 & 2012 Aug 25 & 228 & 28 & 976& $0.66''\times0.46''$ &  190 - 280 & 2011.0.00419.S\\ 
$^{13}$CO (6--5)  & 661.067 & 2012 Dec 01 & 60 & 20 & 1953& $0.49''\times0.25''$ &  110 - 220 & 2011.0.00277.S\\ 
\\[.1mm]
C$^{18}$O (2--1)  & 219.560 & 2014 Apr 04 & 158 & 34 & 244& $1.03''\times0.74''$ &  200 - 250 & 2012.1.00564.S\\ 
  &  & 2014 Apr 10 & 158 & 32 & 488& $0.79''\times0.65''$  & 200 - 250 & 2012.1.00496.S\\
C$^{18}$O (3--2)  & 329.331 & 2012 Jun 02 & 230  & 18 & 488 & $0.52''\times0.46''$ &  225 - 275 & 2011.0.00318.S\\ 
  &  & 2012 Jun 17 & 197 & 21 & 488 & $0.86''\times0.48''$ &  225 - 275 & 2011.0.00318.S\\ 
  &  & 2012 Aug 10 & 198 & 26 & 488 & $0.71''\times0.98''$ &  225 - 275 & 2011.0.00318.S\\  
\\[.1mm]
C$^{17}$O (3--2)  & 337.061 & 2014 Apr 27 & 158 & 31 & 1953& $0.53''\times0.40''$ &  200 - 250 & 2012.1.00453.S\\ 
  &  & 2014 May 27 & 158 & 36 & 1953& $0.43''\times0.35''$ &  200 - 250 & 2012.1.00453.S
\enddata
\tablenotetext{a}{After Hanning smoothing by the correlator, equivalent to twice the channel spacing.}
\tablenotetext{b}{FWHM of the Gaussian restoring beam.}
\tablenotetext{c}{Derived from contribution functions after modeling.}
\end{deluxetable*}

CO is an extremely stable molecule in Titan's atmosphere with a chemical lifetime greater than 75 Myr \citep{kra14}. Additionally, the atmosphere does not reach temperatures cold enough for CO to condense, suggesting that CO is most likely uniformly mixed throughout the atmosphere. Recent measurements of atmospheric CO abundances using various instruments (see Table~1) have converged to a volume mixing ratio (VMR) of approximately 50 ppm with no evidence for altitude variation. 

The derivation of isotopic ratios can provide significant information needed to unravel the complex formation and/or subsequent evolution of Titan based on isotopic deviations from measured values of other bodies in the Solar System \citep{nixon12, mandt12}. Deviations may be a consequence of primordial differences in the isotopic abundances of these molecules or may have emerged over time due to preferential selection of certain isotopes during chemical reactions via the kinetic isotope effect. Recent measurements of the $^{12}$C/$^{13}$C ratio from various species in Titan's atmosphere show little evidence for deviation from the terrestrial standard value of 89.3 \citep{lodders03}. Measurements of the $^{16}$O/$^{18}$O ratio have suggested a possible enrichment in $^{18}$O from the terrestrial value, although these determinations contain significant uncertainty \citep{nixon08b,courtin11,gurwell11}.

The advent of the Atacama Large Millimeter/submillimeter Array (ALMA) provides the ability to probe this region in greater detail with unprecedented spectral and spatial resolution at high sensitivity. Previous ALMA studies of Titan have presented mapping and vertical column densities of hydrogen isocyanide (HNC) and cyanoacetylene (HC$_{3}$N) \citep{cordiner14a}, as well as the first spectroscopic detection of ethyl cyanide (C$_{2}$H$_{5}$CN) in Titan's atmosphere \citep{cordiner15}.

In this study, we present observations of rotational emission lines of CO and its isotopologues in Titan's atmosphere with the highest spectral resolution and sensitivity achieved to date, as well as the first spectroscopic detection of C$^{17}$O in the outer solar system.  C$^{17}$O was expected to be present within the atmosphere, but its emission lines have been too weak to detect until now. This study demonstrates the power of ALMA for future planetary studies and provides a better understanding of the atmospheric evolution and dynamics associated with oxygen-bearing molecules in Titan's atmosphere. 

\begin{figure*}
\centering
\includegraphics[width=0.33\textwidth]{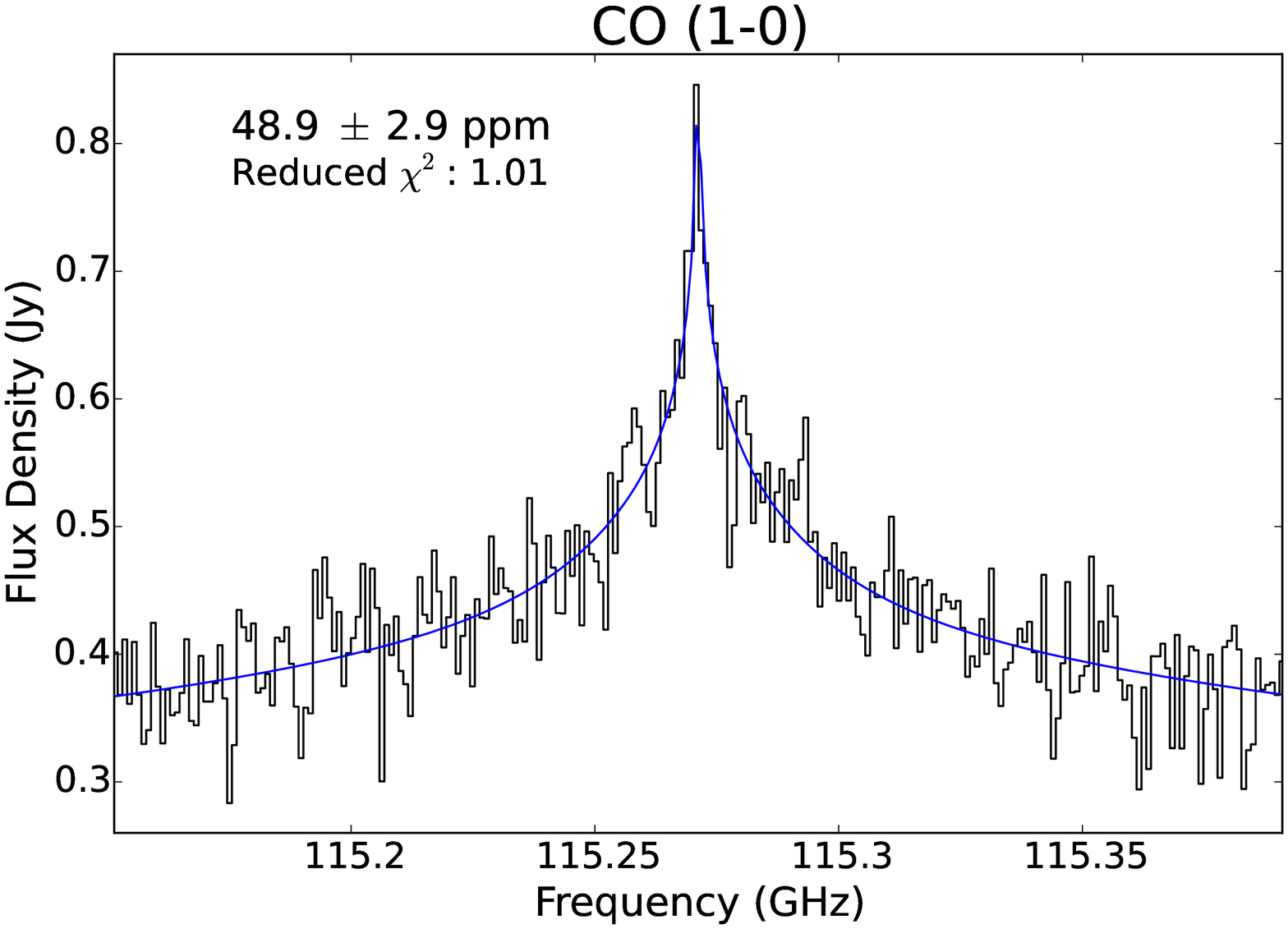}
\includegraphics[width=0.33\textwidth]{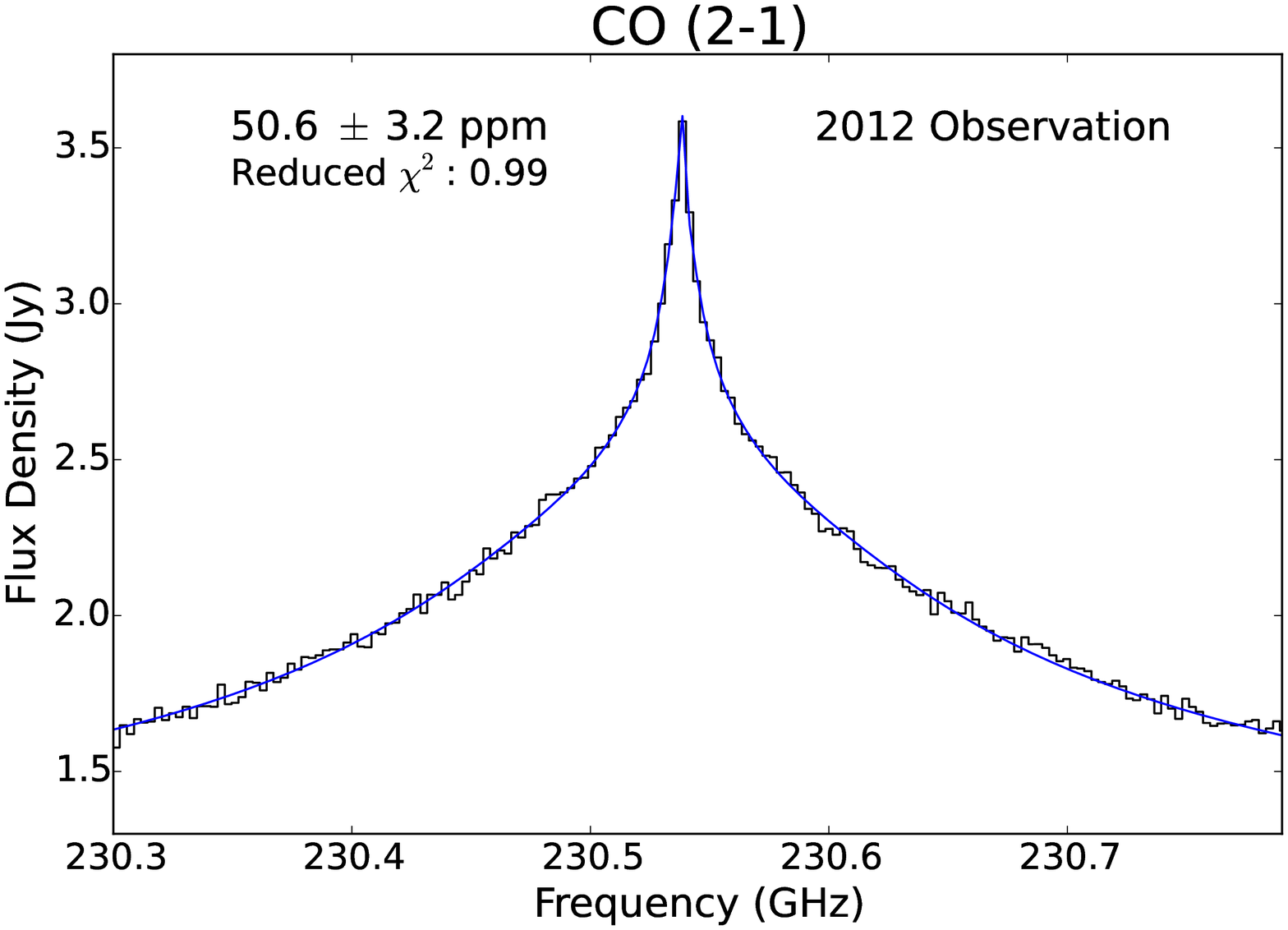}
\includegraphics[width=0.33\textwidth]{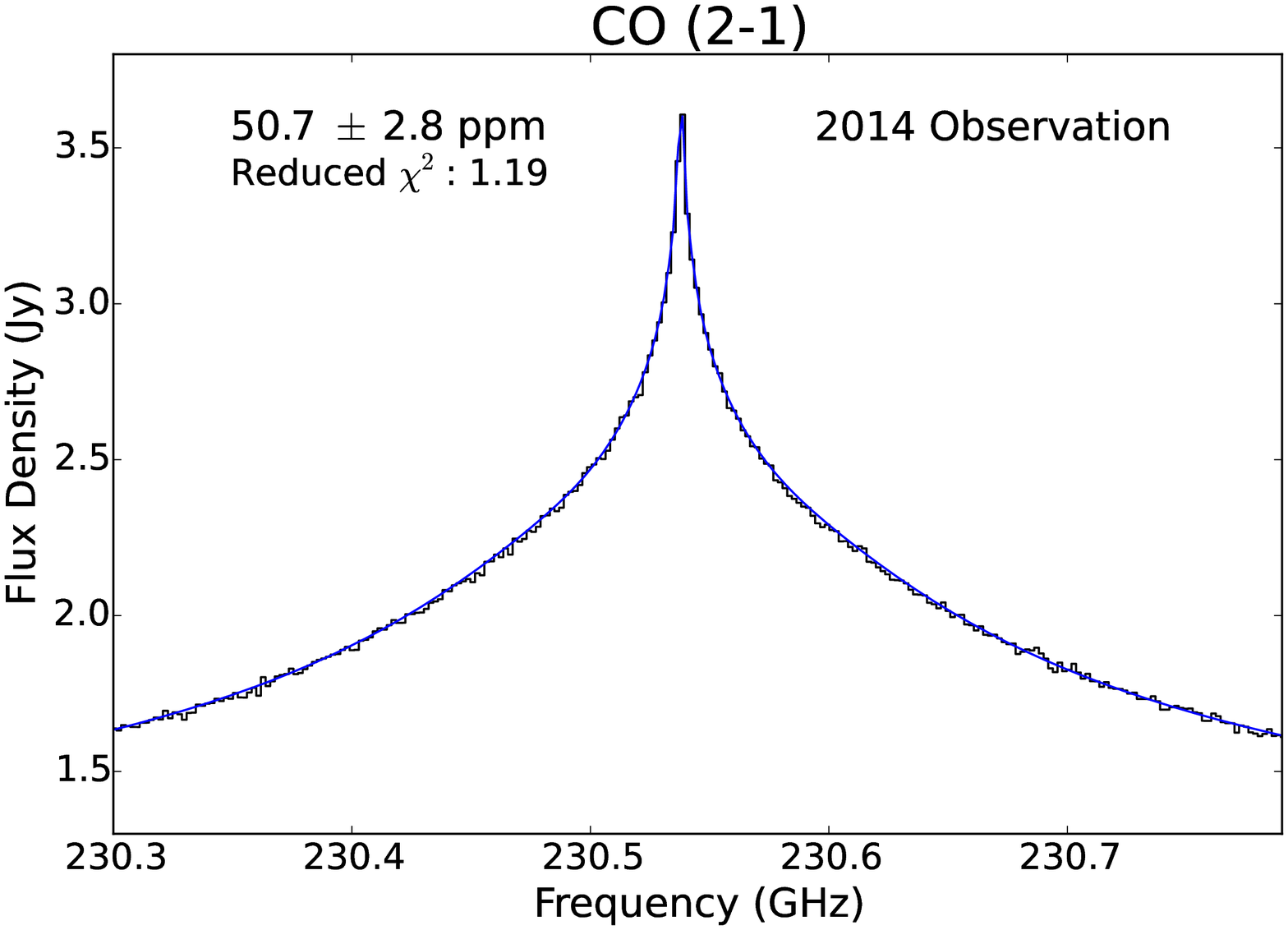}
\includegraphics[width=0.33\textwidth]{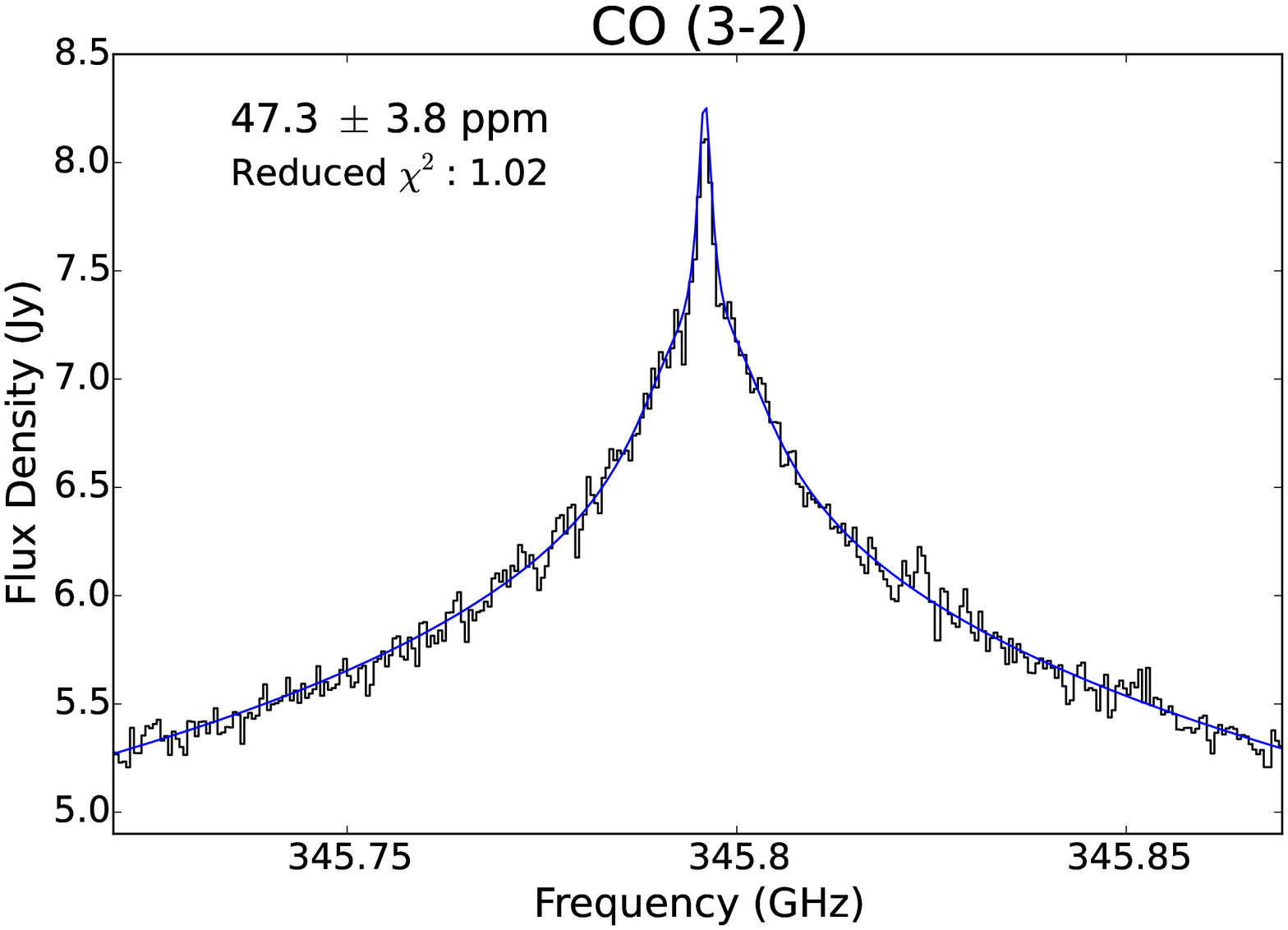}
\includegraphics[width=0.33\textwidth]{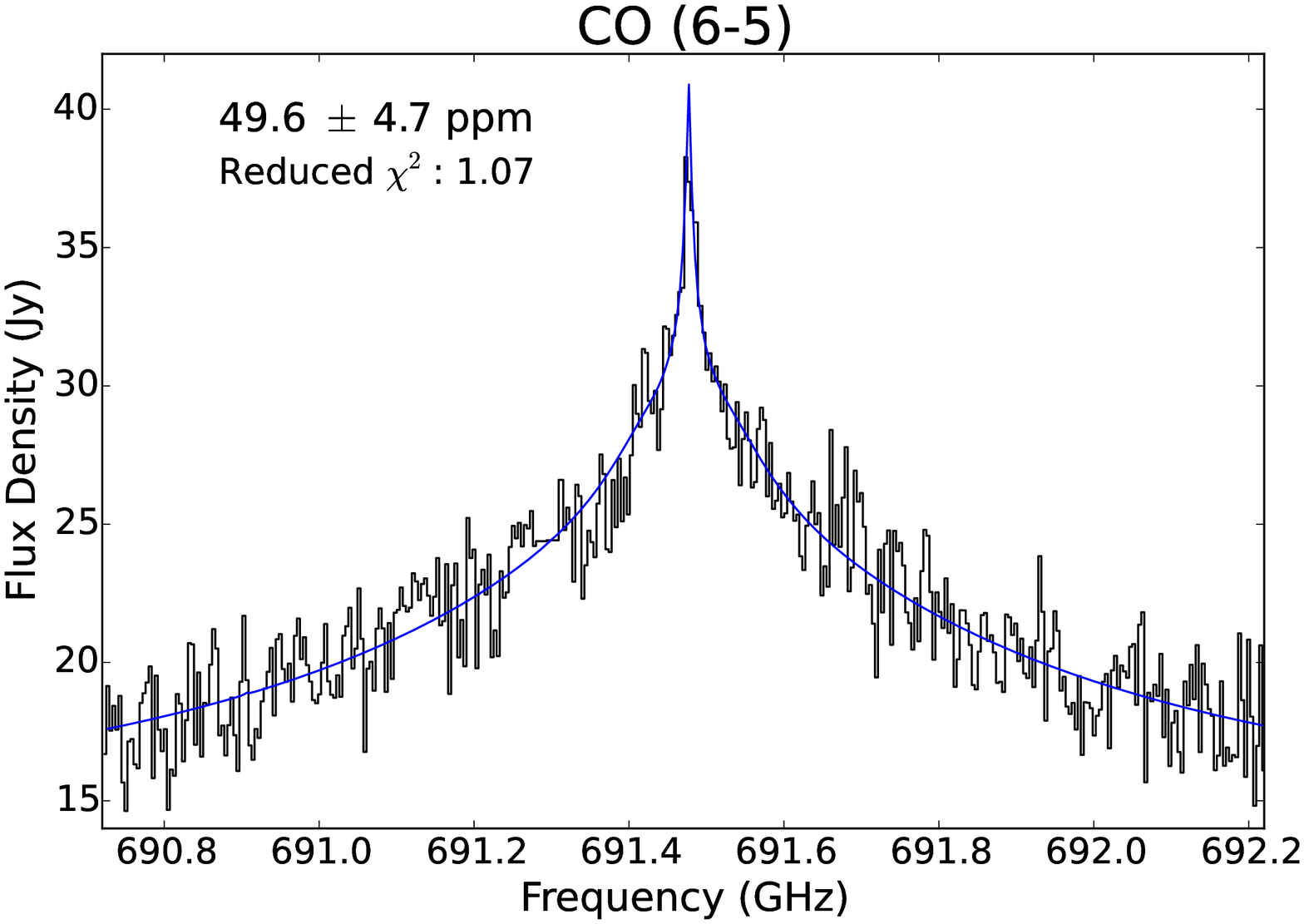}
\includegraphics[width=0.33\textwidth]{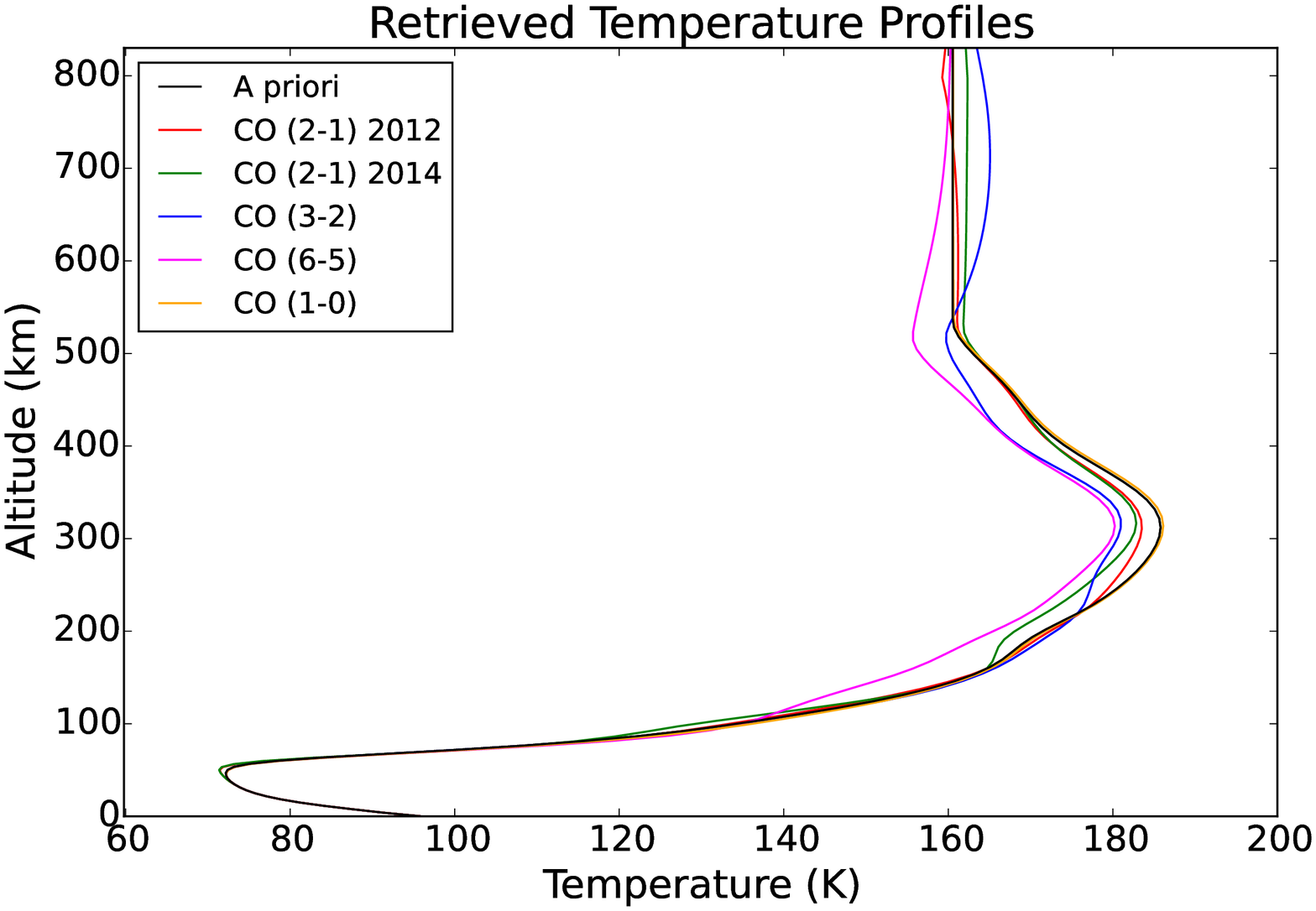}
\caption{Spatially integrated spectra of CO with best-fitting models overlaid. Emission lines were modeled individually and best-fitting abundances from each model (top left of each plot) were combined to achieve a result of 49.6 ${\pm}$ 1.8 ppm. Bottom right: Retrieved temperature profiles for each observation.}
\end{figure*}

\section{Observations}

Interferometric observations of Titan (obtained from the ALMA Science Archive\footnote{https://almascience.nrao.edu/alma-data/archive} and summarized in Table 2) were taken on multiple dates in 2012 and 2014. ALMA often uses Titan to calibrate the absolute flux scale for the science target, resulting in numerous short observations of Titan spanning the lifetime of ALMA. Despite short integration times, high signal-to-noise is achieved due to ALMA's excellent sensitivity.

We selected observation dates within a short period of time to minimize possible effects of seasonal variations in Titan's thermal structure. Data sets from 2012 were a part of ALMA Early Science Cycle 0 observations, when the full array was not yet completed. Our study was therefore expanded into 2014 (ALMA Early Science Cycle 1), which included a more complete array of up to 36 antennae and allowed for improved signal-to-noise in order to measure the less abundant isotopologues of CO more precisely. Observations from 2012 were treated separately from 2014 observations and results were combined after modeling. Data from ALMA Bands 3, 6, 7, and 9 were used for this study.

Titan's diameter ranged from 0.67$^{\prime\prime}$ to 0.78$^{\prime\prime}$ on the sky throughout the observation dates. Since the size of Titan was similar to the ALMA beam size, this study uses disc-averaged spectra. Self-calibration of the phases was possible due to Titan's strong and compact continuum emission. ALMA was set to track Titan's ephemeris position and the coordinates of the phase center were updated in real-time. Weather conditions were good for the observation dates chosen, with precipitable water vapor (PWV) at zenith ranging from 0.46 to 2.87 mm and an average PWV of 1.38 mm. Bright quasars (e.g., 3C279) were used for bandpass calibration.

The data were flagged and calibrated using revised versions of the calibration scripts provided by the Joint ALMA Observatory. Revisions included unflagging desired emission lines and updating the Titan flux model (see ALMA Memo 594\footnote{https://science.nrao.edu/facilities/alma/aboutALMA/Techno-\allowbreak logy/ALMA\textunderscore Memo\textunderscore Series/alma594/memo594.pdf}). The initial flux scale was established by matching the continuum level from a relatively line free region in the spectrum to that of the Butler-JPL-Horizons 2012 flux model for Titan, which is expected to be accurate to within 15\%.

Data reduction used a similar method to \citet{cordiner15} using version 4.3.1 of the Common Astronomy Software Applications package (CASA) \citep{mcm07}. Image deconvolution was performed using the H{\"o}gbom clean algorithm with natural visibility weighting and a threshold flux level of twice the expected RMS noise. The Splatalogue database for astronomical spectroscopy\footnote{http://www.cv.nrao.edu/php/splat/} was used to assign rest frequencies to the observed emission lines and spectral coordinates were corrected into Titan's rest frame using JPL Horizons On-Line Ephemeris System\footnote{http://ssd.jpl.nasa.gov/horizons.cgi}.

A total of 15 detections of 10 different transitions of CO and its isotopologues were used to obtain the results of this paper. Detections are listed in Table~2. Measurements of C$^{18}$O and C$^{17}$O were combined for each transition using the CASA \textit{fixplanets} and \textit{cvel} tasks, increasing the signal-to-noise of these weaker transitions.

\begin{figure*}
\includegraphics[width=0.33\textwidth]{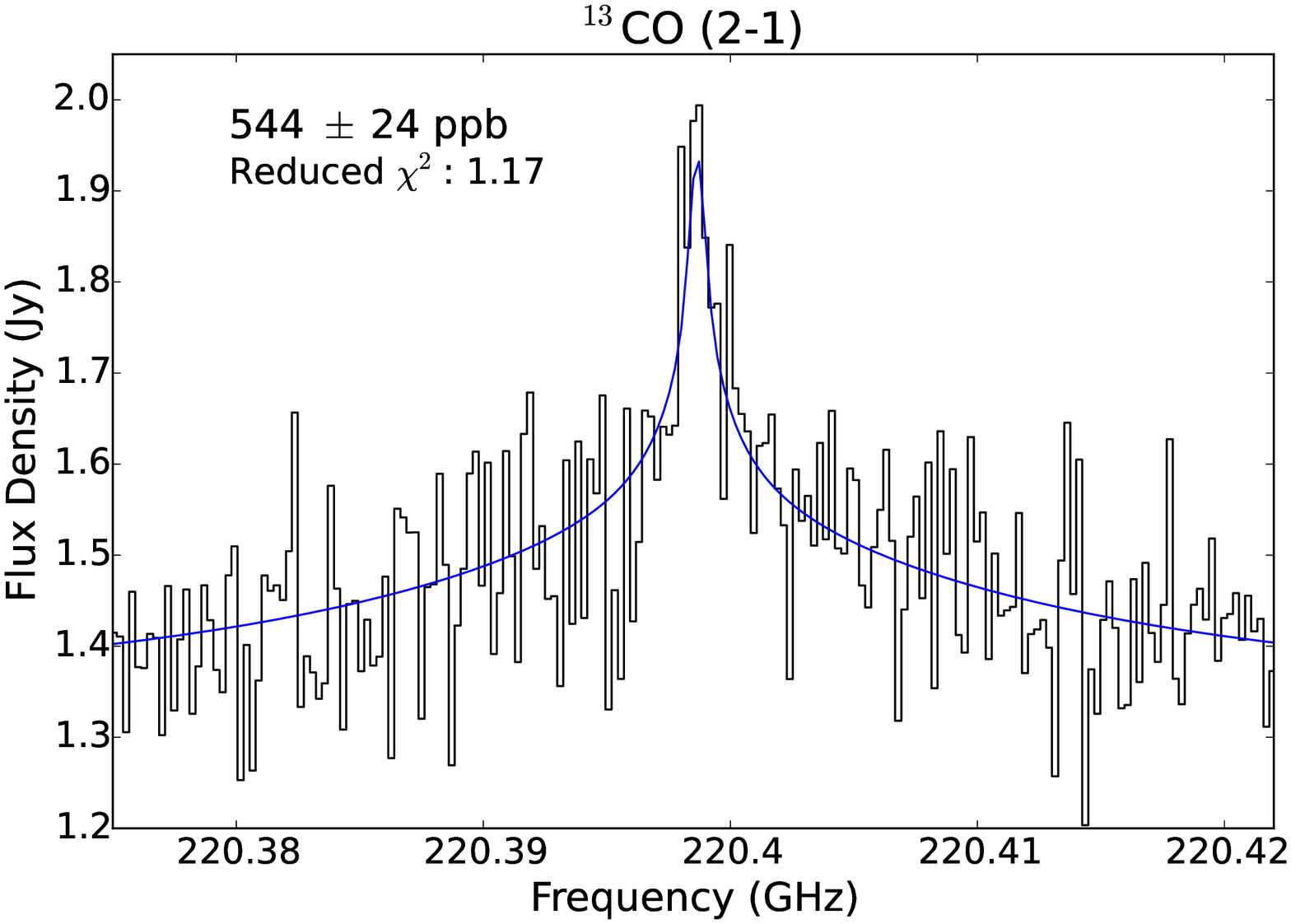}
\includegraphics[width=0.33\textwidth]{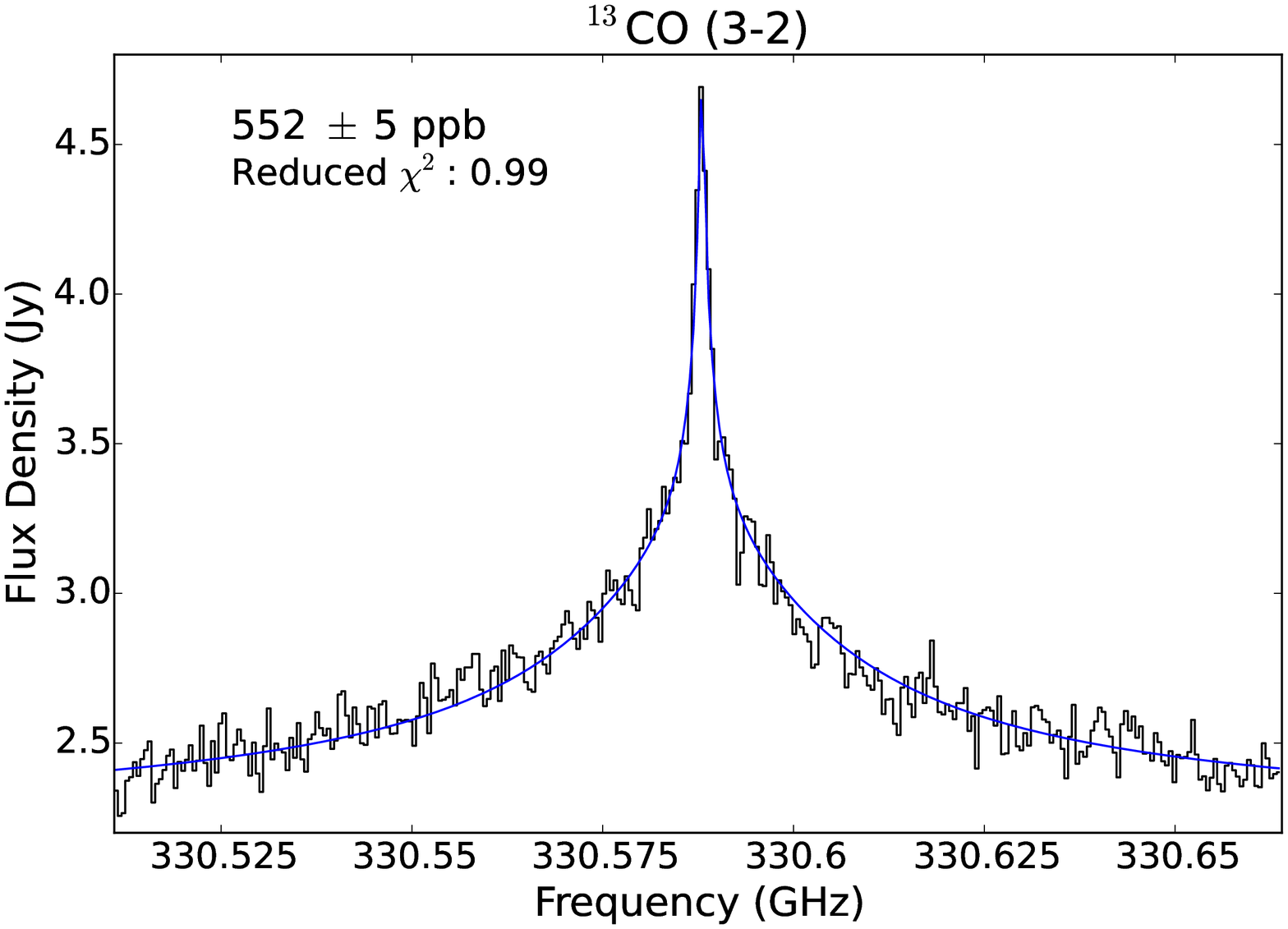}
\includegraphics[width=0.33\textwidth]{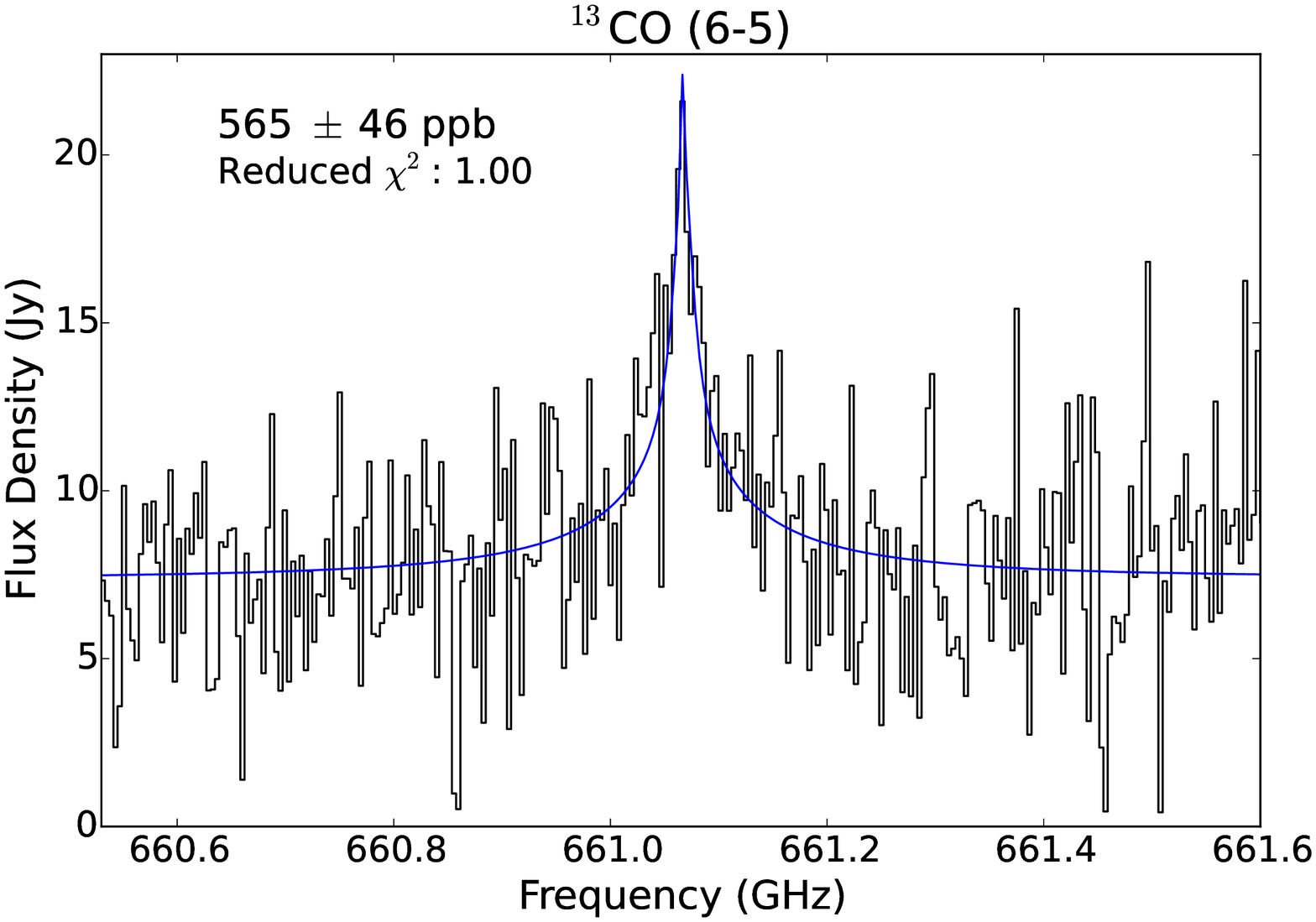}
\includegraphics[width=0.33\textwidth]{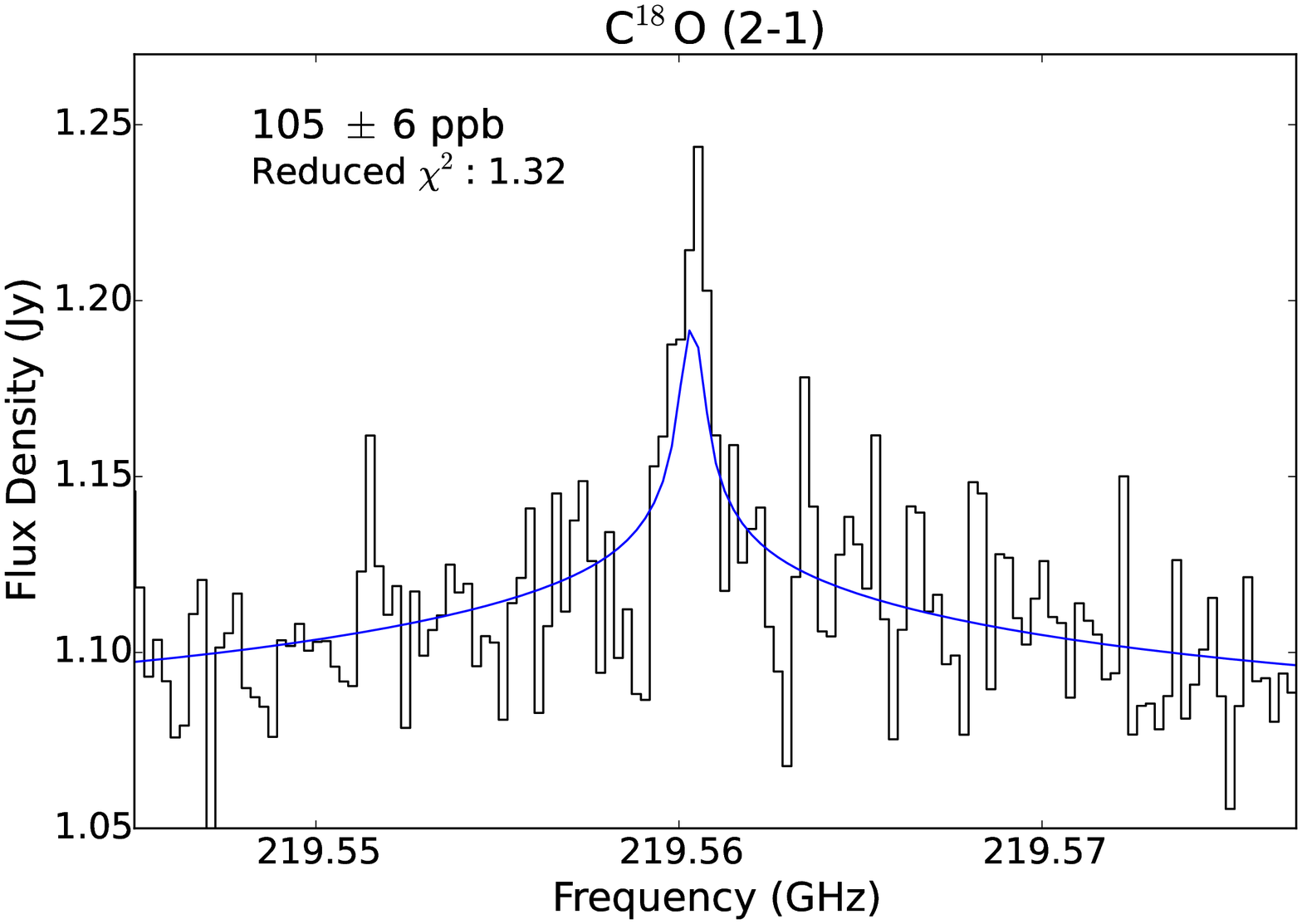}
\includegraphics[width=0.33\textwidth]{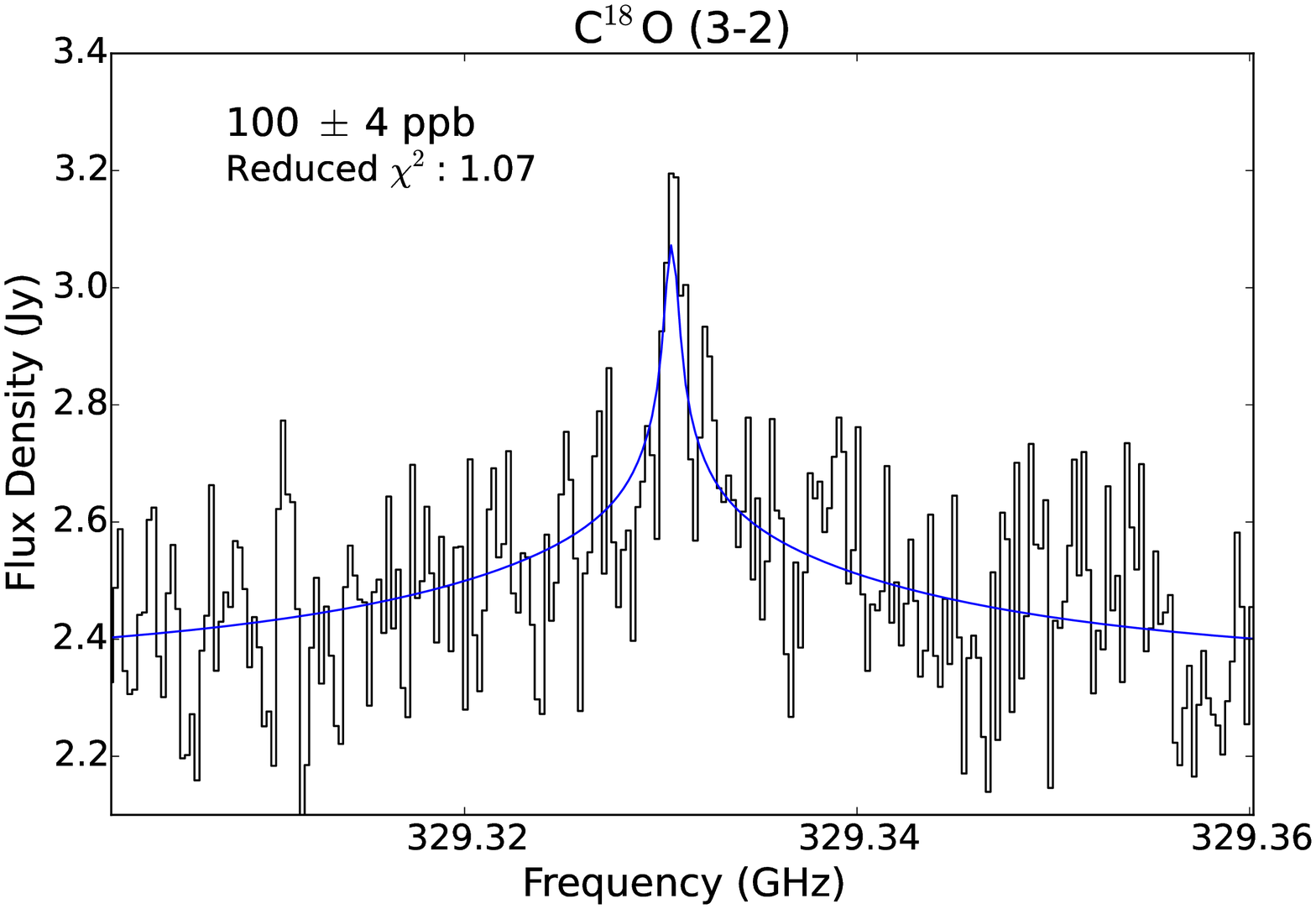}
\includegraphics[width=0.33\textwidth]{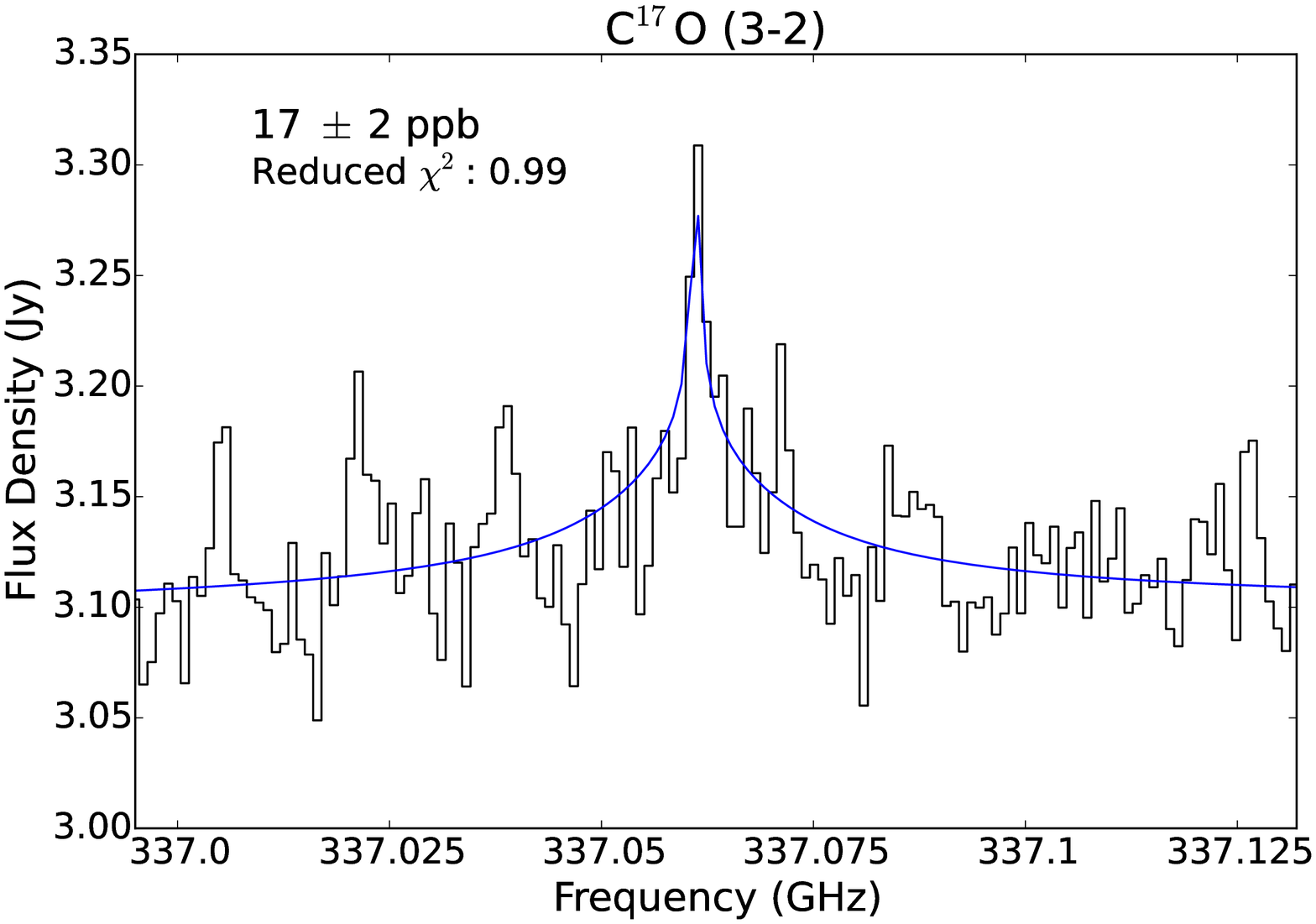}
\caption{Top: Spatially integrated spectra of $^{13}$CO with best-fitting models overlaid. Bottom: Spatially-integrated spectra of C$^{18}$O and C$^{17}$O (3--2) with best-fitting models overlaid. C$^{18}$O (2--1) and C$^{17}$O (3--2) are each combinations of two observations taken close in time to each other, while C$^{18}$O (3--2) is a combination of three observations. Observations were combined in CASA before deconvolution.}
\end{figure*}

\section{Modeling and Results}

Modeled and observed spectra for CO are presented in Fig.~1. We obtained spectra by integrating over a region of the ALMA data cube that includes the flux from Titan's solid disc and atmosphere out to 850 km plus 3$\sigma$$_{PSF}$, where $\sigma$$_{PSF}$ is the standard deviation of the Gaussian restoring beam (point spread function) with FWHM equal to the major axis of the beam ellipse. Although spectra were originally extracted in units of Janskys, the spatial distribution of radiance must be considered in order to enable modeling and account for the variations in Titan's geocentric distance between observations. Thus, fluxes were converted to spectral radiance units (W/cm$^{2}$/sr/cm$^{-1}$) following the procedure outlined in \citet{teanby13}.

In a similar manner to \citet{cordiner15}, individual spectra were modeled using the NEMESIS line-by-line radiative transfer code \citep{irw08}. We adopted an initial atmospheric temperature profile derived from a combination of Cassini CIRS and Huygens Atmospheric Structure Instrument (HASI) measurements \citep{flasar05, fulchignoni05}. Abundances of nitrogen and methane isotopes and aerosols were adopted from Teanby et al. (2013) and line parameters were extracted from the HITRAN 2012 database \citep{rothman13}. Collision-induced absorption coefficients from all combinations of N$_2$, CH$_4$, and H$_2$ were calculated following the routines of \citet{borysow86a, borysow86b, borysow86c, borysow87, borysow91, borysow93}. Gas vertical profiles for these molecules were based on \citet{niemann10}. The ALMA spectra were rescaled to match the continuum level predicted by the NEMESIS model, which typically differed from the Butler-JPL-Horizons 2012 flux model by about 10\%. 

The CO lineshape is significantly pressure-broadened in Titan's atmosphere, making it an excellent indicator of vertical temperature structure and allowing for measurements of abundance. Limb brightening due to Titan's extensive atmosphere is also significant. To account for this, a disc-averaged reference spectrum based on a spatial weighting function and field-of-view averaging points must be created for each CO transition, as outlined in Appendix A of \citet{teanby13}. The number of averaging points for this study ranged from 24 to 27, depending on how extensive the limb-brightening effects of each transition appeared. The circular area covered by these points extended out to a radius of 3425 km from Titan's center, which includes the atmosphere out to 850 km above Titan's surface.

\begin{deluxetable*}{lccccc}
\tablewidth{0pt}
\tablecaption{Best-fitting CO Abundances and Isotopic Ratios}
\tablehead{
\colhead{Reference} &
\colhead{Instrument} &
\colhead{CO abundance (ppm)} &
\colhead{$^{12}$C/$^{13}$C} &
\colhead{$^{16}$O/$^{18}$O} &
\colhead{$^{16}$O/$^{17}$O}}
\citet{courtin11} &Herschel SPIRE &40${\pm}$5 &87${\pm}$6 & 380${\pm}$60 & \nodata\\
\citet{gurwell11} &SMA &51${\pm}$4 &84${\pm}$8 & 472${\pm}$104 & \nodata\\
\citet{rengel14} &Herschel PACS &50${\pm}$2  &124${\pm}$58  & \nodata & \nodata\\
This work & ALMA & 49.6${\pm}$1.8 & 89.9${\pm}$3.4  & 486${\pm}$22 & 2917${\pm}$359\\
Terrestrial values & & & 89.3 & 498.8 & 2680.6
\end{deluxetable*}

After defining reference temperature, aerosol, and abundance profiles, we modeled the CO emission lines under the assumption that the mixing ratio remains constant with altitude up to 850 km. We began with an a priori mixing ratio of 50 ppm and simultaneously retrieved a constant vertical abundance and temperature profile. NEMESIS uses an iterative $\chi^2$ minimization technique that relies on both the quality of the fit to the data and the level of deviation from the a priori setup. As a result of simultaneously retrieving gas abundance and temperature, the uniqueness of fit must be determined using correlation coefficients. The highest correlation between temperature and gas abundance exists at an altitude of 100 km, with an absolute correlation coefficient of 0.5. The temperature in this altitude region, however, is well constrained by Huygens measurements \citep{fulchignoni05} and not permitted to deviate significantly from a priori values. We also tested the uniqueness of fit by holding the CO gas abundance at constant values between 40 and 60 ppm and allowing for the spectra to be fit using only temperature retrievals. We found that any deviation in the CO abundance of ${\pm}$~3 ppm from the a priori value of 50 ppm does not allow an adequate fit to the data and requires a vertical temperature profile that is not consistent with observations from Huygens and CIRS. Thus, we conclude that temperature retrieval alone cannot fully compensate for fixed CO abundance past the normal level of error determined in the NEMESIS model. Furthermore, this proves that we are able to discriminate the temperature profile from the CO abundance when modeling a single CO line. Errors for the globally averaged retrieved temperature profile are of the order of 8 K at the top of the atmosphere (above 500 km) and 4 K at 200 km. Each observation was modeled independently and results for all five emission lines were combined using the technique provided in \citet{nixon08a} Section 4. A systematic error in the CO line intensity of 2\% \citep{rothman13} was also added to the measurement error. The best-fitting abundance for CO was found to be 49.6~${\pm}$~1.8 ppm, which is consistent with and more accurate than previous measurements.

Retrieved vertical temperature profiles from CO models were then used as fixed temperature profiles for modeling of the less abundant CO isotopologues ($^{13}$CO, C$^{18}$O, and C$^{17}$O.) For each isotopologue transition, the retrieved temperature profile which was closest in time to that observation was used as the fixed temperature profile with errors from that temperature retrieval propagated into the isotopologue retrieval. For a priori abundances, we assumed terrestrial isotopic ratios of 89.3 for $^{12}$C/$^{13}$C, 498.8 for $^{16}$O/$^{18}$O, and 2680.6 for $^{16}$O/$^{17}$O \citep{lodders03}.

Best-fitting models of CO isotopologues were found using the same modeling procedures as CO (aside from the fixed temperature profile) and are presented in Fig.~2. The combined best-fitting $^{12}$C/$^{13}$C ratio of 89.9 ${\pm}$ 3.4 is consistent with similar studies of other molecules \citep{courtin11, nixon12, mandt12}, suggesting no significant carbon fractionation occurred during CO formation. Best-fitting ratios of $^{16}$O/$^{18}$O and $^{16}$O/$^{17}$O are 486 ${\pm}$ 22 and 2917 ${\pm}$ 359, respectively. Although the C$^{18}$O (2--1) peak appears slightly under-fitted, the spectral residuals (model minus observation) for this transition indicate an optimal fit given the noise. The mixing ratios for C$^{18}$O, derived independently from the J = 3--2 and J = 2--1 transitions, are consistent within the observational errors.

Additional systematic errors that have not been quantified in the final error estimate could exist. Sources of systematic errors include our assumption that Titan has the same temperature profile at all locations in our disc-averaged calculation, as well as our assumption that the Voigt profile is correct. Additional errors could exist in the line-broadening coefficients and temperature-dependence exponent used in the line database \citep{rothman13}, which could alter the line strength and shape. An accurate estimation of these systematic errors is difficult and beyond the scope of this paper, but these possible sources are mentioned for completeness.

Our results constitute the most accurate determinations of atmospheric abundance and isotopic ratios of CO in Titan's atmosphere to date.  This also marks the first detection of $^{17}$O in the outer solar system, with a  $^{16}$O/$^{17}$O isotopic ratio within 1$\sigma$ of the terrestrial value.

\section{Discussion}

Our result for the CO abundance in Titan's atmosphere of 49.6 ${\pm}$ 1.8 ppm is in good agreement with all other recent measurements presented in Table 1. Observations of CO and its isotopologues fit well with a uniform mixing profile. Additionally, different isotopologues and transitions, which probe different levels of Titan's stratosphere and mesosphere (from 140 km up to 550 km, see Table 2), provide consistent abundances and isotopic ratios similar to terrestrial ratios. Thus, the assumption of a constant CO abundance with altitude appears to be validated.

Direct comparisons for isotopic ratios in CO are found in Table 3. The $^{12}$C/$^{13}$C ratio is measured to be approximately constant throughout the solar system, suggesting that the bulk of material originated from a common source which had an isotopic ratio in carbon of approximately 89 (see review of Solar System measurements in \citealt{niemann10}). Measurements of $^{12}$C/$^{13}$C in multiple species in Titan's atmosphere (e.g., CH$_{4}$, CH$_{3}$D, C$_{2}$H$_{2}$, C$_{2}$H$_{6}$, C$_{4}$H$_{2}$, HCN, HC$_{3}$N, CO, CO$_{2}$), as well as our new measurement of 89.9 ${\pm}$ 3.4 in CO, are in accordance with this value. This includes the in situ measurement of this ratio in methane, Titan's most abundant carbon-bearing species, of 91.1 ${\pm}$ 1.4 using the Huygens Gas Chromatograph Mass Spectrometer (GCMS) \citep{niemann10}. Our result further suggests that no significant carbon fractionation has occurred during the chemical reactions responsible for producing atmospheric CO on Titan.

The fact that oxygen has three stable isotopes is advantageous in studying its chemical evolution. This allows for  discrimination between preserved primordial characteristics of the solar nebula and later perturbations due to chemical reactions occurring within the atmosphere, which have different rates of fractionation. Ratios in oxygen are less consistent throughout the solar system and are not as widely studied as the $^{12}$C/$^{13}$C ratio due to greater difficulty in accurately measuring the abundances of oxygen isotopologues.

\citet{noll95} reported the detection of a single H$_{2}$$^{18}$O line on Jupiter, implying a $^{16}$O/$^{18}$O ratio of 1--3 times the terrestrial value of 498.8. A low signal-to-noise ratio has also been measured in CO$_{2}$ on Titan by \citet{nixon08b} to be 346 ${\pm}$ 110. Previous Titan measurements of $^{16}$O/$^{18}$O in CO include 380 ${\pm}$ 60 by \citet{courtin11} and 472 ${\pm}$ 104 by \citet{gurwell11}. All three measurements on Titan suggest a slight enrichment in $^{18}$O, although the most recent measurement by \citet{gurwell11} is closer to the terrestrial standard. Our result of 486 ${\pm}$ 22 includes the terrestrial value within the 1$\sigma$ error. Oxygen's least abundant stable isotope, $^{17}$O, is difficult to measure due to its low abundance. Consequently, direct comparison of our $^{16}$O/$^{17}$O ratio of 2917~${\pm}$~359 to other outer solar system bodies is not possible since no other $^{17}$O features have been observed. 

Since oxygen is presumed to precipitate into Titan's atmosphere from the Enceladus torus, a study of the oxygen ratios in the plumes of Enceladus could provide insight into the evolution of oxygen on Titan. Furthermore, a better understanding of oxygen processes in Titan's atmosphere can be realized with dedicated observations and longer integration times using ALMA's completed array. These data will include a higher signal-to-noise ratio, allowing for additional detections of transitions of CO's weaker isotopologues and better constraints on the abundances and isotopic ratios. 

Following the end of the Cassini mission in 2017, ALMA and other ground and space-based alternatives will provide the primary means to further our understanding of the evolution and dynamical processes associated with Titan's atmosphere. This study exemplifies the power of ALMA, even during Early Science Mode, as an important tool for studying bodies in our Solar System. At its full capacity, ALMA's capability to continue to search for new molecular detections at higher sensitivity, map latitudinal and longitudinal variations at higher angular resolution, and track seasonal variations on Titan will make it an extremely important tool to continue improving our understanding of this complex atmosphere.

\acknowledgments

We would like to acknowledge the staff at the helpdesk of the North American ALMA Science Center (NAASC) in Charlottesville, Virginia for providing helpful information on the calibration of ALMA data. NAT and PGJI are funded by the UK Science and Technology Facilities Council. JEL is supported by an appointment to the NASA Postdoctoral Program at the NASA Goddard Space Flight Center, administered by Oak Ridge Associated Universities through a contract with NASA. This work makes use of the ADS/JAO.ALMA project codes listed in Table 2. ALMA is a partnership of ESO (representing its member states), NSF (USA) and NINS (Japan), together with NRC (Canada), NSC and ASIAA (Taiwan), and KASI (Republic of Korea), in cooperation with the Republic of Chile. The Joint ALMA Observatory is operated by ESO, AUI/NRAO and NAOJ. The National Radio Astronomy Observatory is a facility of the National Science Foundation operated under cooperative agreement by Associated Universities, Inc.

\end{document}